\documentclass[aps,prd,twocolumn,10pt,amsmath,reprint,groupedaddress,amssymb,amsfonts,nofootinbib,longbibliography]{revtex4-1}

\usepackage{graphicx}
\usepackage{amssymb}
\usepackage{amsmath}
\usepackage{longtable}
\usepackage[utf8]{inputenc} 
\usepackage{color}
\usepackage{supertabular}
\usepackage{dcolumn}
\usepackage{ wasysym }
\usepackage{subfigure}
\usepackage{notoccite}

\def\DMoff2F{2\mathcal{F}_\textrm{DM-off}}
\def\DMon2F{2\mathcal{F}_\textrm{DM-on}}

\newcommand{\avgSeg}[1]{\overline{#1}}			

\newcommand{\Freq}{f}
\newcommand{\fdot}{{\dot{\Freq}}}

\newcommand{\F}{\mathcal{F}}		

\newcommand{\avF}{\avgSeg{\F}}

\let\svthefootnote\thefootnote

\begin{document}
\date{\today}
\title{A new veto for continuous gravitational wave searches}
\author{Sylvia J. Zhu$^\mathrm{1,2,a}$, Maria Alessandra Papa$^\mathrm{1,2,3,b}$, Sin\'{e}ad Walsh$^\mathrm{3}$\\\vspace{0.3in}
}\let\thefootnote\relax\footnote{\textsuperscript{a}email: sylvia.zhu@aei.mpg.de}\let\thefootnote\relax\footnote{\textsuperscript{b}email: maria.alessandra.papa@aei.mpg.de}
\affiliation{$^1$ Max-Planck-Institut f{\"u}r Gravitationsphysik, am M{\"u}hlenberg 1, 14476, Potsdam-Golm\\
$^2$ Max-Planck-Institut f{\"u}r Gravitationsphysik, Callinstra{$\beta$}e 38, 30167, Hannover\\
$^3$ University of Wisconsin-Milwaukee, Milwaukee, Wisconsin 53201, USA\\
\vspace{0.1in}}
\addtocounter{footnote}{-2}\let\thefootnote\svthefootnote

\begin{abstract}
  We present a new veto procedure to distinguish between continuous
  gravitational wave (CW) signals and the detector artifacts that can mimic
  their behavior. The veto procedure exploits the fact that a long-lasting coherent 
  disturbance is less likely than a real signal to exhibit a Doppler modulation of astrophysical 
  origin. 
  Therefore, in the presence of an outlier from a search, we perform a multi-step 
  search around the frequency of the outlier with the Doppler modulation turned off (DM-off),
  and compare these results with the results from the original (DM-on) search. If the results from the DM-off search are 
  more significant than those from the DM-on search, the outlier is most likely due to an artifact rather than a signal. 
  We tune the veto procedure so that it has a very low false dismissal rate. 
  With this veto, we are able to identify as coherent disturbances $>\!99.9\%$ of the 6349 candidates from the
  recent all-sky low-frequency Einstein@Home search on the data from the Advanced LIGO
  O1 observing run \cite{O1AS20-100}. We present the details of each identified disturbance
  in the Appendix. 
  \end{abstract}

\preprint{LIGO-P}
\maketitle

\section{Introduction}
\label{sec:introduction}

In searches for continuous gravitational waves (CWs) from rotating neutron
stars with asymmetries, detector artifacts can partly mimic
the behavior of astrophysical signals and yield high values of the detection
statistic, so that these artifacts look more like signals than Gaussian noise.
In recent years, detection statistics that are more robust to
lines (the line-robust statistic \cite{Keitel2014}) and
  transient disturbances (the line-and-transient-robust statistic
  \cite{Keitel:2016}) have been developed.
However, in spite of the fact that these exhibit higher detection efficiencies in
the presence of the disturbances that they were designed to be robust against, 
searches that have used such statistics can still suffer from a
large number of loud candidates caused by detector artifacts \cite{O1AS20-100}.

A standard CW signal is approximately monochromatic with a phase
evolution that is characterized by a frequency $f_0$ and its time-derivatives,
e.g., $\fdot_0$. The frequency $f$ at which a detector on Earth receives the signal
is Doppler shifted over time 
with respect to $f_0$ as the Earth rotates and orbits around the Sun. 

Conceptually, when searching for a signal from a given sky location,
the data analysis techniques correct the data for the Doppler shift  
and ``reconcentrate'' all the signal power in a narrow
frequency range --- optimally, a single frequency bin --- at $f_0$. 

Coherent detector artifacts of terrestrial origin do not exhibit this
Doppler modulation but can still match a signal wave-shape well enough
to yield a significant value of the detection statistic in a search for astrophysical signals. However, 
in the results of a search for waveforms without Doppler modulation (DM-off),
the significance of an artifact should increase with respect to the results of an astrophysical search whereas the
significance of an astrophysical signal should decrease. We can use this
difference in behaviors to set up a veto. 

The standard CW waveforms (DM-on) are derived from astrophysical modeling of
neutron stars and their general form is well constrained.
In contrast, detector artifacts can take many forms. Some artifacts have
known causes,
and if their existence is known ahead of time, data from the affected frequency ranges can be
and often are replaced with Gaussian noise (or ``cleaned''; e.g., \cite{S5GC1HF,S6Bucket,O1AS20-100}). 
However, many artifacts only become apparent after a search has been performed; this is especially
true when new data are analyzed.


It is impossible to quantify the performance of a veto that tests for detector disturbances 
other than by using data containing the coherent disturbances 
whose detrimental impact we want to mitigate. In its current form, this veto was developed for the
Einstein@Home all-sky low frequency search of the data from the first observing run of
Advanced LIGO (O1) \cite{O1AS20-100}. The veto characterization and exploration were
performed with this search in mind, so many specific aspects of it (e.g., grid spacings,
frequency band) reflect the \cite{O1AS20-100} search setup. However, the general principles
are not search dependent, and this veto can easily be modified to accommodate other
search setups.

This paper is organized as follows: Section \ref{sec:characterization} describes the veto procedure;
Section \ref{sec:O1AS20-100} demonstrates its effectiveness in a real-world application;
and Section \ref{sec:Conclusion} summarizes and discusses results and prospects.

\section{Veto characterization}
\label{sec:characterization}

\subsection{Detection statistic}
\label{sec:detectionStat}

This method can be used to test candidates from any continuous wave search.
However, we concentrate here on Einstein@Home-like searches, which have typically
included the deepest investigations \cite{S6BucketFU}.

Einstein@Home searches for continuous waves typically search over $10^{17}$ different
templates and return only the top candidate per million. These results must then be analyzed
with automated methods that rely on the data being reasonably well behaved, so all frequency
bands that contain visible large-scale patterns 
are set aside and not analyzed upfront (see for example \cite{S6CasA}). Hence, the remaining data are fairly ``clean.''
  The line-and-transient-robust statistic \cite{Keitel2014,Keitel:2016} is used to 
  select the candidates that are most likely to be signals. An estimate of their significance against Gaussian noise fluctuations is performed  
  using the standard average $2\avF$ statistic (where the average is taken over the different
segments of the semi-coherent search) \cite{JKS, Cutler+Schutz}.

In this paper, we concentrate on applying the veto to candidates
that have survived a whole hierarchy of follow-ups, the last stage of which is a
fully coherent search over the entire observation time \cite{S6BucketFU,O1AS20-100}. 
In this case, the reference detection statistic is $2\F$ rather than the average $2\avF$. 
In stationary Gaussian noise, the $2\F$ statistic follows a central $\chi^2$ distribution 
with four degrees of freedom ($\chi^2_4$). If a signal is present, the observed distribution is
a noncentral $\chi^2$ distribution with four degrees of freedom and a noncentrality
parameter $\rho^2$ that depends on the strength of the signal ($\chi^2_4(\rho^2)$) \cite{JKS}. 



\subsection{The DM-off waveform}

The heart of the DM-off veto is the comparison between the values of the detection
statistic with an astrophysical waveform template bank and the values of the detection
statistic from waveforms without Doppler modulation. The astrophysical waveforms
not only present a frequency modulation due to the Doppler shift 
but also an amplitude modulation.
This is due to the fact that interferometeric detectors 
have non-uniform antenna sensitivity patterns
across the sky and hence that the detector response to the same signal changes in time as the Earth rotates. 
The amplitude modulation depends on waveform parameters that are not explicitly searched for
(hence called the nuisance parameters), because the $\F$-statistic is analytically maximised over them.
We keep the amplitude modulation in the DM-off template waveforms for two reasons: The
detection statistic remains a $\chi^2$ distribution with 4 degrees of freedom, which
allows for a straightforward comparison with the results of the astrophysical search; and, the
amplitude of the disturbance waveform is allowed to vary and potentially match
  a larger variety of disturbance types than a fixed amplitude modulation would allow.

\begin{figure}[h]
  \includegraphics[width=\columnwidth]{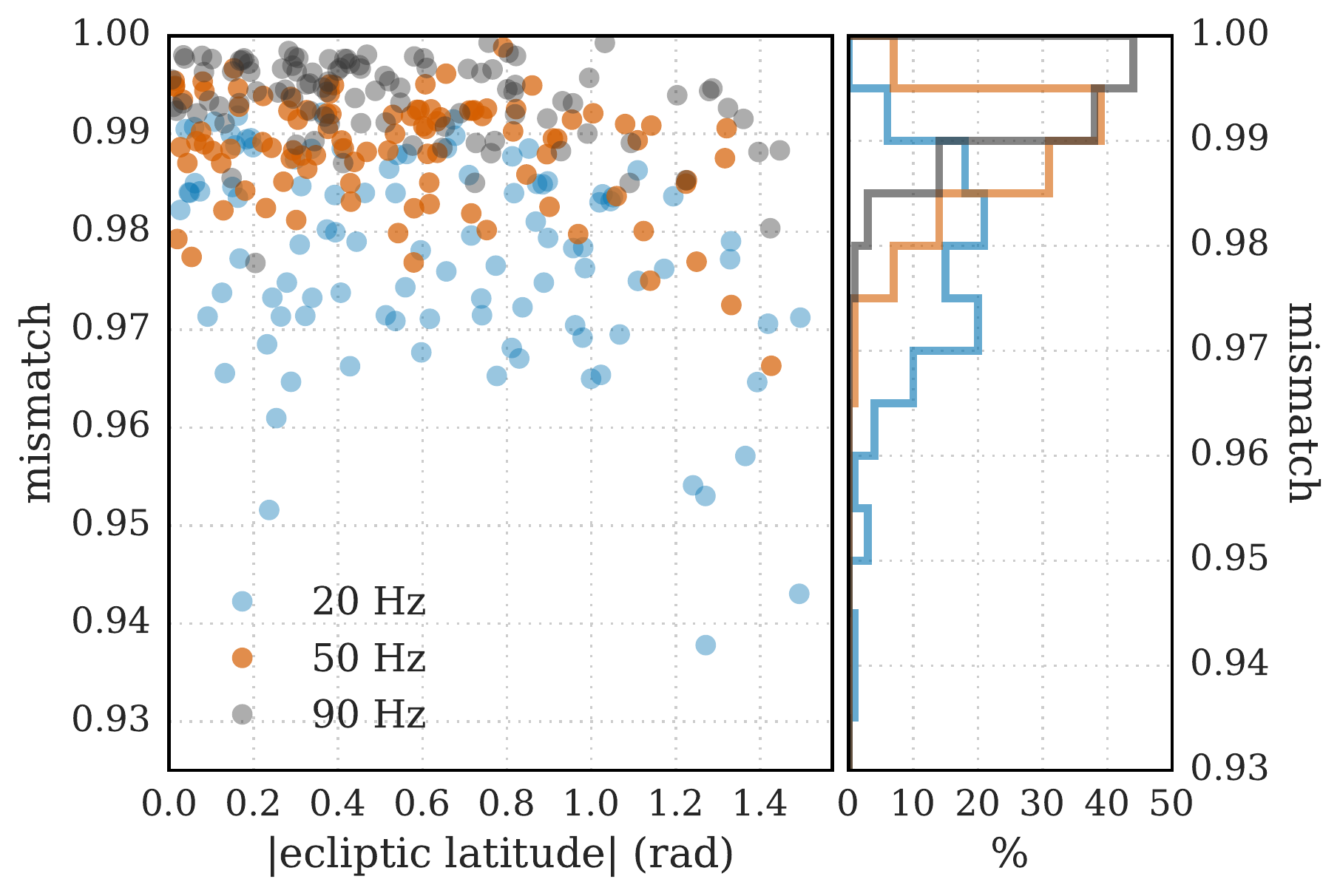}
  \caption{The mismatch gives a measure of the dissimilarity between the DM-off waveforms
    and an astrophysical signal. The left panel here shows the mismatch as a function
    of the latitude of the source, where zero is at the equator
    and $\pi/2$ is at the poles. The right panel has the histograms for those
    mismatches. There is a dependence 
    on the latitude and on frequency of the signal, as one would expect;
      sources at lower ecliptic latitudes and with higher frequencies have
  larger mismatches.}
  \label{fig:mismatch}
\end{figure}

How similar are the DM-on and DM-off template waveforms? To answer this question,
we simulate 300 signals for an observation time and with a duty factor
similar to that of the Advanced LIGO O1 observing run. 
The signal frequencies lie in the following ranges: 20-25~Hz, 50-55~Hz, and 90-95~Hz;
the sky positions of the source are random in the sky;
the cosine of the inclination angle is uniformly distributed; 
and the data has no noise. We search each signal using a fully coherent
$\F$-statistic search with a search setup like that of the last follow-up stage of \cite{O1AS20-100}.
We use two template banks: For the DM-on search we use the standard astrophysical
template waveforms, and for the DM-off search (our veto search) we use DM-off template waveforms.
For the DM-on search, we use a single template that is perfectly matched to the signal.
For the DM-off search, we use a grid of templates defined by frequency, spindown, and
sky position, with the grid spacings informed by \cite{O1AS20-100}; see Section~\ref{sec:setup} for details.
From each search we consider the highest detection statistic value: 
$\DMon2F=\rho^2_{\textrm{DM-on}}$ 
and $\DMoff2F=\rho^2_{\textrm{DM-off}}$.
We then compute the mismatch $\mu$:
\begin{equation}
\mu={{\rho^2_{\textrm{DM-on}}-\rho^2_{\textrm{DM-off}}}\over{\rho^2_{\textrm{DM-on}}}}.
\label{eq:mismatch}
\end{equation}
%
Fig.~\ref{fig:mismatch} shows this mismatch at the three frequency ranges.
As expected, higher mismatches are found for signals that
experience larger Doppler shifts. This occurs at the ecliptic equator (where
the relative motion between the Earth and the source is greatest)
and at higher frequencies (as the difference between the Doppler-shifted
frequency and the source frequency is proportional to the source frequency). 
We can see that these waveform families are very different: the average mismatch is very high ($>\!98\%$) for even the low frequency signals at 20~Hz.

We expect the mismatch to increase for search setups with longer ($>$ several months)
observation times because the Doppler signature is even stronger for such signals. 
O1 had a particularly short duration,
so DM-off waveforms corresponding to future LIGO observing runs will, in general, have
even larger mismatches with the astrophysical signals.
\subsection{Proof of principle}
\label{sec:S6BucketFU}

We illustrate the concept of this veto by comparing the $\DMon2F$ and the
$\DMoff2F$ of two of the ten candidates that survived the penultimate stage
of a search for continuous wave signals on data from the last Initial LIGO
run (S6) \cite{S6BucketFU}. We consider candidate 1 and 6 of Table II of \cite{S6BucketFU}.
\begin{figure}[h]
  \includegraphics[width=\columnwidth]{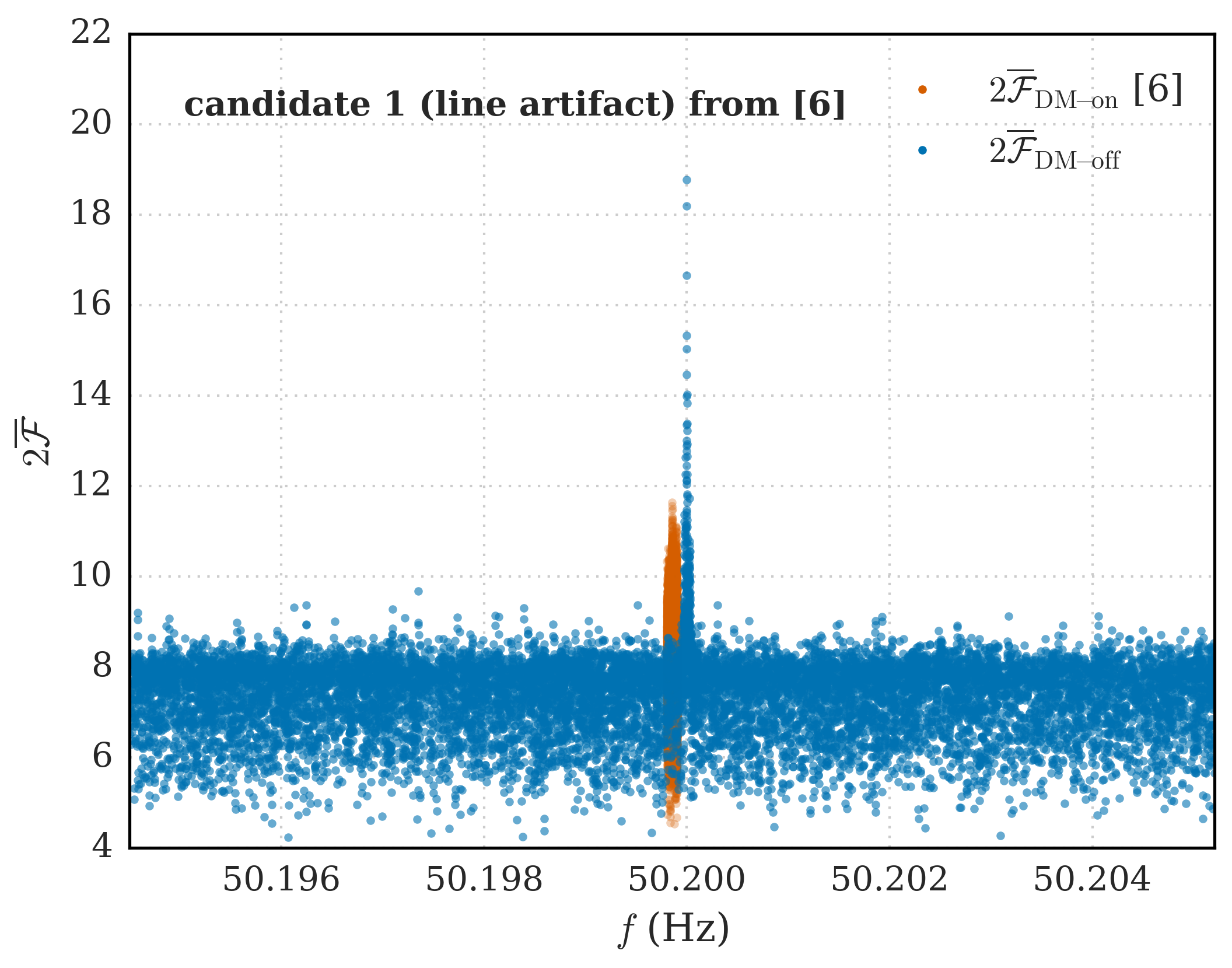}
  \includegraphics[width=\columnwidth]{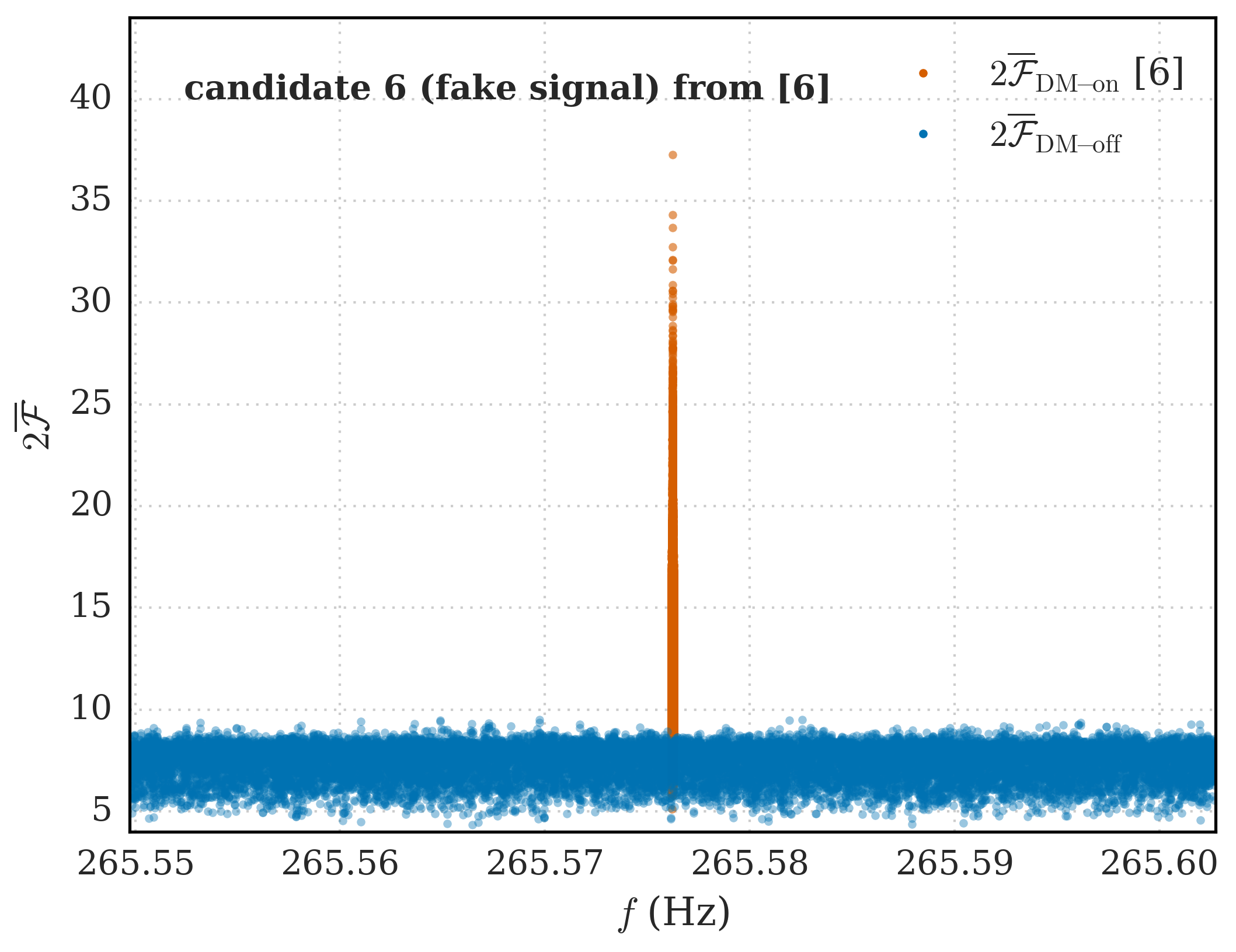}
  \caption{Comparison between the original search results (orange) and the DM-off
    search results (blue) for candidates 1 and 6 of \cite{S6BucketFU}.  Candidate 1 (top)
    did not pass the detection threshold set at $2\avF=15$ in \cite{S6BucketFU};
    candidate 6 (bottom) did, but was associated with a fake continuous wave signal in the data.
    After a DM-off search, the line that caused candidate 1 becomes
    more prominent while the signal from candidate 6 disappears
    into the noise. We use these two candidates to illustrate that
  the DM-off veto concept works.}
  \label{fig:S6Bucket}
\end{figure}
After the last stage candidate 1 did not exceed the $2\avF$  
detection threshold and further studies found an association with a comb of lines.  In contrast,
the detection statistic value for candidate 6 did exceed the detection threshold; however,
candidate 6 was due to a fake signal that was present in the data for validation purposes.  

Figure \ref{fig:S6Bucket} shows the original search results and the DM-off search results
for these two candidates. The comparison is striking: the $2\avF$ statistic associated
with candidate 1 in a DM-off search is a factor of 1.6 larger than the values from the
original search, clearly showing that a non-astrophysical waveform matches the data
better than the astrophysical one. The opposite happens for the candidate associated
with a fake signal: when the data is searched with non-astrophysical waveforms, the
significance of the candidate disappears and becomes comparable with the neighboring noise floor.

\subsection{Implementation}
\label{sec:setup}

In the previous section, we have always referred to the multi-detector $2\F$
(or average $2\avF$
) statistic. This statistic coherently combines the data from
all available detectors in order to determine whether the data are more likely to
contain Gaussian noise or a waveform. However, high values of the detection statistic
can result when the data contain some coherent disturbance --- even only in one detector ---
that looks more like a signal than Gaussian noise. The multi-detector DM-off search
finds the DM-off waveform that best matches the data consistently across the detectors;
if a coherent disturbance occurs only in one of them, it also attempts to take into account
the data from the other detector and may yield a detection statistic value that is not as
significant as the single detector value and also not significant enough to exceed a
veto threshold. However, if the data contain a signal, both the multi-detector and
single-detector $\DMon2F$ values will be larger than the corresponding $\DMoff2F$ ones.
Therefore, we perform the $\DMon2F$-$\DMoff2F$ comparison not only between the
multi-detector statistics, but also between the single-detector ones. For a candidate to pass
the veto, we require that all the DM-off $2\F$ statistics --- both multi-detector and single-detector
--- successfully pass the thresholds.

The aim of the DM-off search is to check whether it is possible to produce a more
significant detection statistic value from a non-astrophysical template bank
using the same data (a very small frequency portion of the entire data set,
on the order of mHz) that
produced a significant candidate in the astrophysical-signal search. If this
is the case, then the original candidate is discarded. The starting point is
the DM-on search candidate, defined by its parameters 
$(\alpha^\ast,\delta^\ast,\Freq^\ast,\fdot^\ast)^{\textrm{DM-on}}$.
Consider a DM-on search that explored the whole sky, a broad frequency band,
and a spindown-range $(\fdot_{\textrm{min}},\fdot_{\textrm{max}})$ with
typically $|\fdot_{\textrm{max}}| \ll |\fdot_{\textrm{min}}|$ ($\fdot_\textrm{min} < 0$). The DM-off search
has to include the DM-off waveforms with the highest overlap with the candidate
waveform $(\alpha^\ast,\delta^\ast,\Freq^\ast,\fdot^\ast)^{\textrm{DM-on}}$.
A conservative choice (meaning a choice that will include more waveforms than necessary,
but will not exclude any) is to take the following: 
\begin{itemize}
\item a frequency range $\Delta f$ around $\Freq^\ast$ equal to
  $\Delta f_\mathrm{(spindown)} + \Delta f_\mathrm{(Doppler)}$, where
    $\Delta f_\mathrm{(spindown)} = \fdot^\ast T_{obs}$ is the maximum spindown
  frequency shift over the entire observation time and $\Delta f_\mathrm{(Doppler)}$
  is the Doppler shift freqency during the observation time from a source at the
  sky position ($\alpha^\ast, \delta^\ast$) of the candidate;
\item a broader spindown range than that of the original search: $(\fdot_{\textrm{min}},|\fdot_{\textrm{min}}|)$, where $\dot{f}_\textrm{min} < 0$;
\item the whole sky. 
\end{itemize}

There is a natural scale for template bank spacings of $\Freq$ and $\fdot$
for a DM-off search over an observation time $T_{\textrm{obs}}$:
$1\over {T_{\textrm{obs}}}$ and $1\over {T_{\textrm{obs}}^2}$, respectively.
These are the smallest differences in $\Freq$ and $\fdot$ that
would be resolvable in a sinusoidal signal search over an observation duration $T_{\textrm{obs}}$.
With the generous search ranges given above, such resolutions can result in
high computing costs. On the other hand, the resolution in the sky is only moderate;
it is determined by the 
detectors' antenna patterns (amplitude modulation functions; see Eqs.~(12) and (13) of \cite{JKS}) and 
simple tests show that 
O(50) points in the sky suffice to provide adequate sky resolution for 
arteficts to be recovered by DM-off waveforms.

In a standard search, the grid spacings must be sufficiently fine in order
to minimize the possibility of missing a signal; that is, the goal of a DM-on
search is to find signal candidates. In contrast, the goal
of a DM-off search is to \emph{remove} noise-candidates, so coarser grids merely
reduce the veto's power to identify artifacts. We
can overcome this limitation by applying a finer DM-off grid in a later
step, after we have eliminated most of the candidates, and in this way save computing cycles.

We have designed the DM-off veto in 3 steps. The first two steps
use a search grid in 
$\Freq$ and $\fdot$ that is $10\times 10$ times coarser than the DM-on search grid.
In all the DM-off steps we use a sky grid that comprises
45 sky points isotropically distributed on the celestial sphere.

\begin{enumerate}
\item \textbf{Step 1: Coarse grid, multi-detector.} A multi-detector DM-off search
  around each candidate is run using coarse $\Freq$ and $\fdot$ grids. 
The largest value of $\DMoff2F$ is considered and for a 
candidate to pass to the next stage its detection statistic value must stay below
a predetermined threshold. Such a threshold is set using simulations of searches 
on fake signals to ensure that
  no astrophysical signal would be rejected; see Section~\ref{sec:thresholds} for
details.
\item \textbf{Step 2: Coarse grid, single-detector.} Since instrumental artifacts
  are not expected to be coherent across detectors, the DM-off search is then
  run on the data from the detectors separately on the surviving candidates from Step 1.
  A candidate's $\DMoff2F$ values must be lower than the predetermined thresholds in \emph{all}
  the detectors separately in order to not be vetoed; that is, if a candidate looks like an
  artifact in any detector, then it is discarded. 
\item \textbf{Step 3: Fine grid, multi-detector and single-detector.} In this stage, the $\Freq$ and
  $\fdot$ grid spacings are the same as the grids from the original search, and
  typically this means that $\Freq$ and $\fdot$ are over-resolved. Assuming that
  the location in parameter space of the local maximum of $\DMoff2F$ for the
  candidate has been identified in the previous stages, this stage is a refinement
  in its immediate neighborhood. 4000x4000 fine-grid points in frequency and
  spindown are explored, with both a multi-detector search and single-detector searches.
  The center point for each of these searches is the maximum that was recorded in the previous steps. 
  If a candidate's $\DMoff2F$ value is above the predetermined
  threshold in any of these three fine grid searches, then the candidate is discarded.
\end{enumerate}

\subsection{Thresholds and veto safety}
\label{sec:thresholds}

We concentrate now on the thresholds of the veto in the case of candidates stemming
from a fully coherent search like the last stage (FU3) of \cite{O1AS20-100};
i.e., a 2-detector search on approximately 4 months of data. These studies would
need to be repeated for candidates from a different search, but the general
concept and the effectiveness of the procedure are well illustrated even in a particular case, like this one.

To determine the veto thresholds, we consider a population of 1500 fake signals in
artificial Gaussian noise. We perform both a DM-on search and a DM-off search and
compare the respective detection statistic values. We set the thresholds so that the
veto is safe; i.e., no signal is discarded. We ensure that this happens at all the steps
listed in Section~\ref{sec:setup}.

Most of the fake signals we search have frequencies around 20~Hz, but we include
400 fake signals at 50~Hz and 90~Hz as well to check for consistency across frequencies.
This reflects the population of FU3 candidates from \cite{O1AS20-100},
which are more abundant at lower frequencies. The $\DMon2F$ values
of our fake signals vary between a few (consistent with noise)
to over ten thousand; i.e., we sample the expected behavior
for a wide range of signal strengths. Realistically, we expect any CW signal we
detect to be on the lower end of this range, as otherwise they would have already
been found by a previous search.

The DM-off search parameter space and grids were described in the previous Section.

\begin{figure}
  \includegraphics[width=\columnwidth]{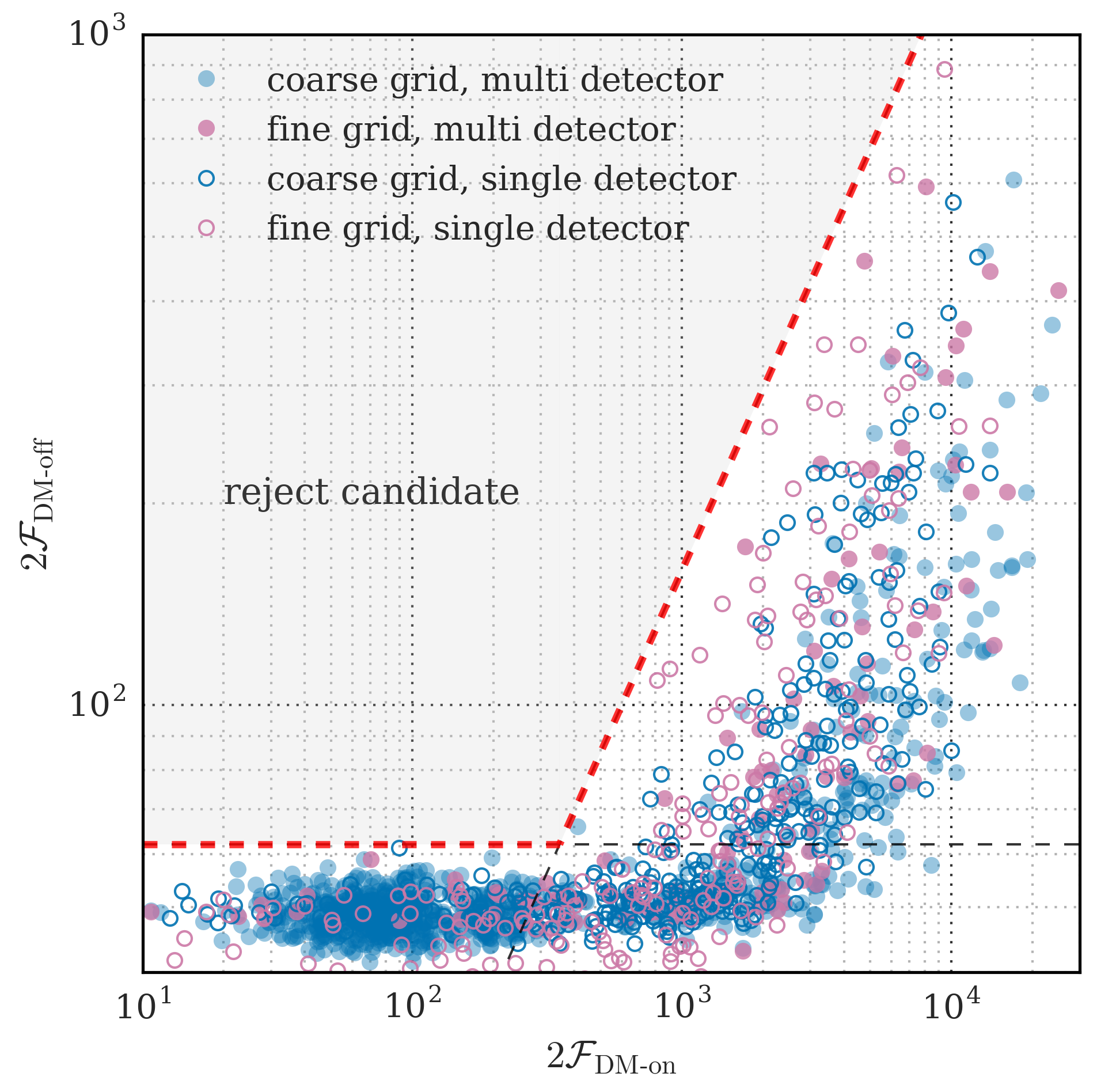}
  \caption{Data containing fake noise plus 1500 simulated signals are searched with both the DM-off
    and the DM-on searches. The highest detection statistics are recorded as $\DMoff2F$
    and $\DMon2F$, respectively. For signals with $\DMon2F \lesssim 400$, the $\DMoff2F$
    is not distinguishable from the noise. The loudest value of
    $\DMoff2F$ found for this group is 61.25. For louder $\DMon2F$, 
    the $\DMoff2F$ values are significantly above the noise and increase with
    $\DMon2F$. The veto threshold is indicated by the red dashed line,
    and was chosen so that none of the fake signals would have been rejected hence yielding a false
  dismissal rate of astrophysical signals smaller than one in 1500.
    Candidates that fall in the region above the red dashed line do not behave like signals,
    and therefore can be rejected.}
  \label{fig:threshold}
\end{figure}

Figure~\ref{fig:threshold} shows the $\DMoff2F$ vs $\DMon2F$ values for 
the 1500 fake signals. For values of $\DMon2F \lesssim 400$, the
$\DMoff2F$ value is always less than 62 (the largest observed $\DMoff2F$ is
61.25) and independent of the $\DMon2F$. In fact, the $\DMoff2F$ values for these weaker
signals are consistent with noise: When
the Doppler modulation is turned off, the detection statistic value is lower than or
comparable to the highest expected value over as many independent trials as there are
templates in the bank, simply due to noise fluctuations. 

For each injection, we determine the $\DMon2F$ value by running an FU3 astrophysical
search on the simulated data in a neighborhood of the signal parameters and taking the
loudest resulting detection statistic value.

Our simulation includes 
all of the steps in the DM-off veto described in Section~\ref{sec:setup}.
Since the results from the different DM-off searches do not differ significantly from each other, we use the same thresholds for all steps (Figure~\ref{fig:threshold}).

For candidates with
$\DMon2F \le 400$, we set a flat threshold at 62; that is, we reject any candidate
at any step with $\DMoff2F \ge 62$. For candidates with $\DMon2F > 400$, we
set a threshold based on their $\DMon2F$ values such that none of our injections
would have been rejected. The threshold, then, is defined as the following:
\begin{equation}
  2\F^\mathrm{thr}_\textrm{DM-off} =
  \begin{cases}
    62 & \DMon2F \le 400,\\
    10^{0.9\log(\DMon2F - 0.5)} & \DMon2F > 400.
  \end{cases}
\label{eq:thresholds}
\end{equation}
We reject a candidate if $\DMoff2F > 2\F^\mathrm{thr}_\textrm{DM-off}.$ 
This threshold is
indicated in Figure~\ref{fig:threshold} by the dashed red line. Based on
our simulations, this threshold yields a false dismissal rate of less than one in 1500, or $\!<0.07\%$.


\section{Effectiveness of this veto with a worked example}
\label{sec:O1AS20-100}

In order to illustrate the effectiveness of this veto, we now show
its noise rejection performance on real detector data. We cannot resort to
synthetic data because only real data has the type of coherent disturbances
that make this veto necessary in the first place, and these populations of
disturbances are not characterized. Therefore, in this Section, we illustrate
the application of the veto on the results of a low frequency search for
continuous signals on Advanced LIGO data \cite{O1AS20-100}.

Compared to previous runs, the low frequency range of the data in this
first Advanced LIGO run was plagued by many coherent disturbances. In
spite of all the ``cleaning'' efforts aimed at identifying disturbed spectral
regions up front, a large excess of detection statistic values above
the pre-set threshold was found. A hierarchical procedure consisting of
a clustering algorithm \cite{clustering} and three follow-up searches reduced
the number of candidates from over 15 million to several thousand. In the end,
6349 signal candidates survived the final fully coherent multi-detector follow-up
stage (FU3), as well as 201 candidates stemming from a fake signal
  present in the data for pipeline-validation purposes.
With the veto that we present in this paper, we are able to
discard the overwhelming majority of the 6349 signal candidates (6345 out of 6349).

The 6349 candidates are not distributed uniformly across the
20-100 Hz frequency range, but are instead clustered in 57
frequency bands 
\cite{O1AS20-100}.
The other 201 candidates all have frequencies at 52.8~Hz,
the frequency of the fake signal. We keep these candidates and use them as
an additional verification of the veto's safety.


\subsection{Veto application}

\begin{table}
  \centering
  \begin{tabular}{|c|c|c|}
    \hline \hline
    Stage       &  num surviving    &  num remaining  \\
                &  candidates        &  frequency bands \\
    \hline \hline
    DM-on Stage 3  & 6349 (201)      & 57 (1) \\
    DM-off Step 1       & 653 (8)         & 22 (1) \\
    DM-off Step 2       & 101 (5)         & 10 (1) \\
    DM-off Step 3       & 4   (1)         & 4  (1) \\
    \hline \hline
  \end{tabular}
  \caption{The results presented in \cite{O1AS20-100} use the DM-off veto
    to test the 6349 candidates that survived 
   multiple stages of an astrophysical search. 
    In order to illustrate the performance of the veto, we present the details of
    that application. The numbers in parentheses denote the number of candidates
    or frequency bands attributed to a fake signal at 52.8~Hz.
    After applying the three steps of the
    DM-off veto, only four candidates survive (as well as a fifth
    candidate due to the fake signal at 52.8~Hz).}
  \label{table:summary}
\end{table}

After {\bf{Step 1}}, 90\% of the initial 6349 candidates are rejected as being
consistent with instrumental artifacts. 653 candidates from 22
frequency ranges pass Step 1 (and 8 from the fake signal, all from the same frequency band).

\begin{figure}
  \includegraphics[width=\columnwidth]{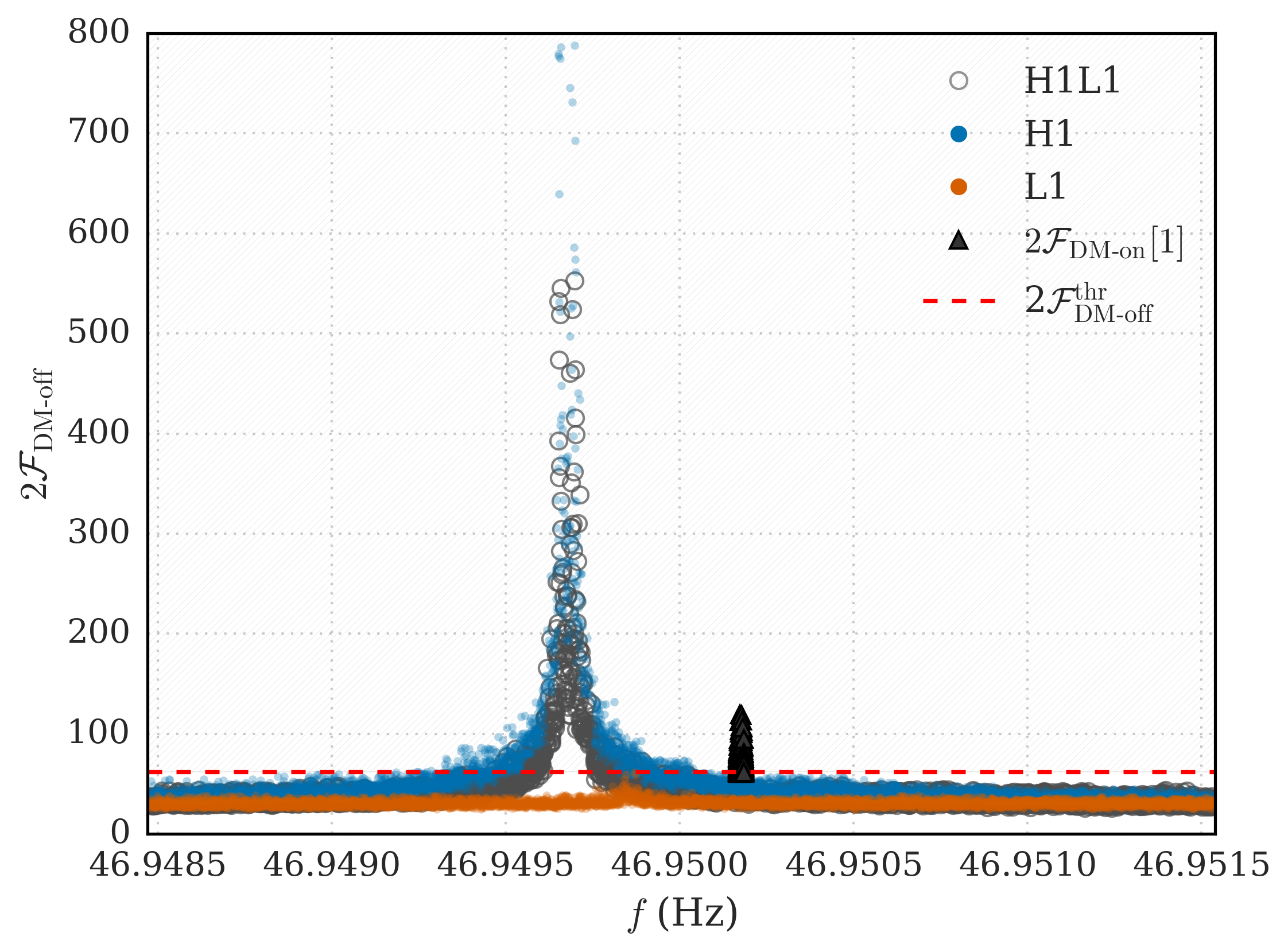}
\caption{An example of a candidate that does not survive Step 1. For
    this candidate, $\DMoff2F$ in H1 exceeds the threshold (the horizontal
  red dashed line) and so this candidate is discarded. This particular
  candidate's elevated $\DMon2F$ value is caused by a stationary line in H1.}
  \label{example_step1_killed}
\end{figure}

Figure~\ref{example_step1_killed} shows an example of a candidate
that does not survive Step 1. Based on the DM-off search results,
we understand that this particular candidate, like all the others
in its frequency region, is caused by a stationary line in H1. The candidate has a value
of $\DMon2F = 80$ at the end of FU3; based on our simulations, signals that
would produce $\DMon2F$ values such as this should have $\DMoff2F < 62$;
in contrast, for this candidate $\DMoff2F \approx 560$ in H1, and so it is discarded.
We note that the DM-off result for L1 is well below the threshold.

\begin{figure}
  \includegraphics[width=\columnwidth]{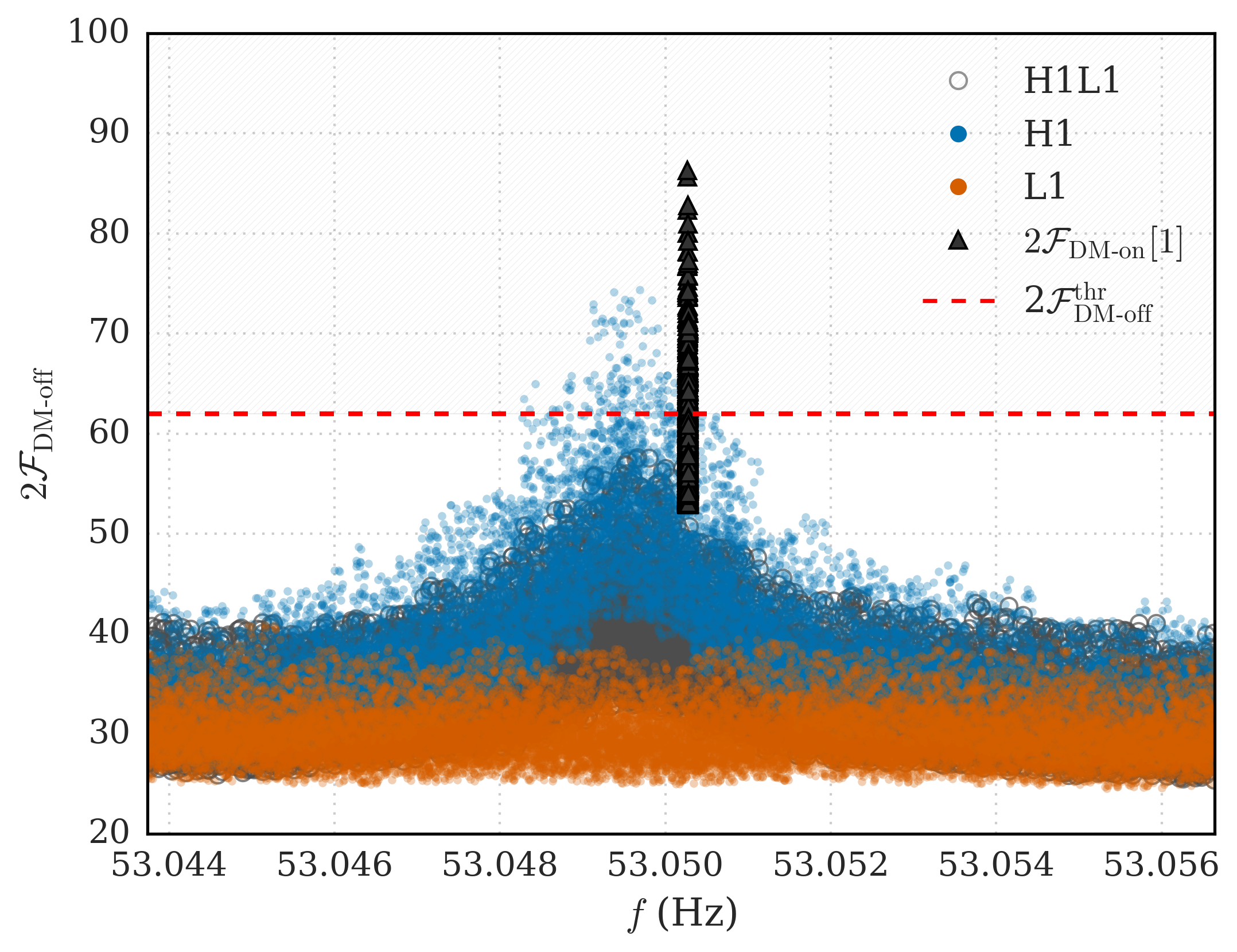}
\caption{An example of a candidate that does not survive Step 2.
    This candidate comes from a wandering line in H1. }
\label{example_step2_killed}
\end{figure}

After {\bf{Step 2}}, 85\% of the remaining 653 candidates are rejected, and
101 candidates from 10 frequency ranges (plus 5 candidates from
the same frequency range due to the fake signal) pass to the next and final step.
Figure~\ref{example_step2_killed} shows an example of a candidate that survives
Step 1 but not Step 2. For this candidate, the H1 $\DMoff2F$ value
exceeds the threshold, but the L1 and the combined H1-L1 $\DMoff2F$ values do not.
In this particular example, the original $\DMon2F$ value is larger than the 
the highest $\DMoff2F$ value; however, the candidate falls in the rejection region (Figure~\ref{fig:threshold}). 
Since the highest $\DMoff2F$ value is obtained using only the H1 data (blue dots) 
and from a template with $\fdot=1.4\times 10^{-10}$ Hz/s, we conclude that this
candidate's elevated $\DMon2F$ value is due to a wandering line (because of the non-zero
spin-up of the DM-off template) in H1. Candidates such as this one illustrate the importance
of requiring that a candidate's single detector and multi-detector
DM-off statistics must all pass the thresholds, and of not restricting
the waveform model to stationary lines (i.e., $\dot{f} \ne 0$).

\begin{figure}
  \includegraphics[width=0.9\columnwidth]{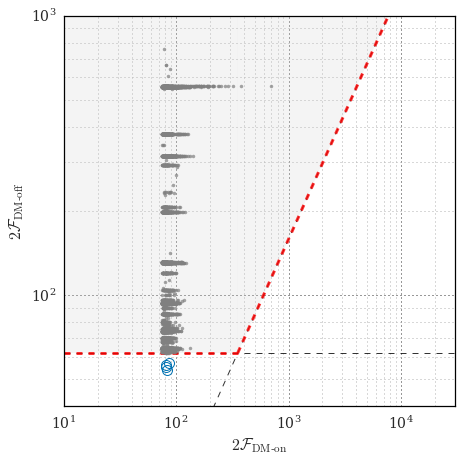}
  \caption{$\DMoff2F$ versus $\DMon2F$ for the 6349 candidates of the example
    that we illustrate in Section \ref{sec:O1AS20-100}. The dashed line
    shows the veto threshold of Eq.~\ref{eq:thresholds}, and the pink and blue
    points are described in Figure~\ref{fig:threshold}. The region above the
    dashed line is the candidate-rejection region. Four signal candidates
    (blue circles) survive the DM-off veto, as well as fifth candidate that is
    due to a fake signal. The remaining 6345 candidates (gray dots) are rejected
    as being caused by detector artifacts.
  }
  \label{resultsSummary}
\end{figure}

Only four candidates pass {\bf{Step 3}} (Figure~\ref{resultsSummary}),
plus one from the fake signal. These four candidates come from different frequency ranges. 

Although this veto is optimized for computational efficiency, its cost is not negligible. 
Its application to the 6349 candidates used $\approx$ 6000 CPU-cores $\times$ 4
hours on nodes of the Atlas cluster \cite{Atlas}. (As a comparison,
  the original search on Einstein@Home cost was divided into 1.9 million work-units with an average
  computational cost of 8 CPU hours per work-unit \cite{O1AS20-100}.)

\section{Conclusion}
\label{sec:Conclusion}

The basic idea behind the DM-off veto is very simple: 
How likely is it that a candidate is due to 
a continuous {\it astrophysical} signal at a certain frequency 
compared to
a continuous {\it terrestrial} signal in the same frequency region?
As we have shown, we can simply ``turn off'' the Doppler modulation of the astrophysical signal
and construct a family of coherent waveforms that has small overlaps with the astrophysical signals
and models well the coherent disturbances in the data.

Based on this, we have developed a simple but highly effective veto to
identify coherent disturbances among candidates of coherent continuous wave searches.
The veto thresholds must be tuned depending on the coherent search set-up with simulations of the veto performance on astrophysical signals, and we show how to do this with a particular example in Section~\ref{sec:thresholds}. 

When this veto is applied to the 6349 candidates from the last and fully coherent stage
of a hierarchical search \cite{O1AS20-100}, it was able to identify the overwhelming
majority of these ($>\!99.9\%$) as being due to noise artifacts. The results of the search,
including the veto, are presented in \cite{O1AS20-100}. In this paper, we show examples
from this veto application, in order to illustrate the method.

The veto is implemented in a hierarchy of three steps: if a candidate is discarded at step 1 (or 2)
it is not subject to step 2 and (or) 3. This reduces the computational burden of this veto,
which, however, still requires an average of $\approx$ 4 CPU-core hours per candidate.

After applying this veto, we not only discard spurious candidates in the astrophysical
searches but also know more about the weak coherent artifacts in the data. We know
where they lie in frequency, what detector is affected, and whether the line is stationary
or wandering (and, through the $\fdot_{\textrm{DM-off}}$, we have a measure of how much it moves).
For instance, half of the 57 frequency ranges that produced the 6349 signal candidates
contain stationary lines. In the Appendix, we provide a Table that characterizes the artifacts,
based on this veto.

It may be possible in the future to use a modified version of this veto in order to identify
stationary lines before an astrophysical search, and to treat these
regions differently or even preemptively exclude them from the analysis.

Not all hierarchical searches for continuous gravitational wave signals have a
final fully coherent stage; see, for instance, \cite{S6BucketFU}. In \cite{S6BucketFU},
no candidates survived Stage 4, which was a 22 segment (280 hrs coherent observation time)
semi-coherent search\footnote{We note that in Section~\ref{sec:S6BucketFU} 
we ran the DM-off searches using the search setup of the last stage of \cite{S6BucketFU}, which is a semi-coherent search, 
so we in fact have used a semi-coherent version of the DM-off veto to illustrate its concept. We are therefore 
confident that a more rigorous generalization will be possible.}. In this context, in a forthcoming paper, we will explore at
what stage the semi-coherent version of this veto might be more effective than the
standard hierarchy of semi-coherent follow-ups. 

In this paper we consider continuous signals from isolated neutron stars. However, the DM-off veto can be
easily generalized to test candidates from other continuous sources, including neutron stars in binary systems.

\acknowledgements
The DM-off veto procedure was used in \cite{O1AS20-100}, and we thank Sergey Klimenko and Evan Goetz for the review of the application of this new veto to the results of that search. We also thank Andrew Melatos, David Keitel, Greg Ashton and Grant Meadors for useful comments. M A Papa and S Walsh gratefully acknowledge the support from NSF PHY Grant 1104902. All computational work for this search was carried out on the ATLAS super-computing cluster at the Max-Planck-Institut f{\"u}r Gravitationsphysik, Hannover and Leibniz Universit{\"a}t Hannover. The authors thank to the LIGO Scientific Collaboration for access to the data and gratefully acknowledge the support of the United States National Science Foundation (NSF) for the construction and operation of the LIGO Laboratory and Advanced LIGO as well as the Science and Technology Facilities Council (STFC) of the United Kingdom, and the Max-Planck-Society (MPS) for support of the construction of Advanced LIGO. Additional support for Advanced LIGO was provided by the Australian Research Council. This document has LIGO ({\texttt{https://dcc.ligo.org}}) DCC number P1700114. \\ \\

\bibliography{bibtexStyle}



\onecolumngrid

\newpage
\appendix
\subsection{Table of artifacts}
\begin{table*}
\begin{tabular}{|c|rcl||c|rcl||c|}
\hline
\hline
\textbf{H1 $f$ (Hz)}  & \multicolumn{3}{c||}{\textbf{H1 $\fdot$ (Hz/s)}} &  \textbf{L1 $f$ (Hz)}  &  \multicolumn{3}{c||}{\textbf{L1 $\fdot$ (Hz/s)}} & comments \\
\hline
\hline
20.404  & 2.172 & $\times$ & $10^{-11}$   &  ---  && --- &&  \\
---     && --- &                          & 21.459   & $-7.036$ & $\times$ & $10^{-13}$ & \\
26.176  & -1.259 & $\times$ & $10^{-10}$  & --- && --- && broad feature\\
26.309  & 1.051 & $\times$ & $10^{-11}$   & --- && --- && loud stationary line \\
---     && --- &                          & 26.343 & $1.600$ & $\times$ & $10^{-10}$ & \\
26.525  & 1.144 & $\times$ & $10^{-11}$   & --- && --- && \\
---     && --- &                          & 27.487   & $-7.263$ & $\times$ & $10^{-11}$ & \\
27.591  & 4.150 & $\times$ & $10^{-10}$   & --- && --- && \\
---     && --- &                          & 27.842  & $-1.638$ & $\times$ & $10^{-12}$ & \\
---     && --- &                          & 27.895  & $-6.890$ & $\times$ & $10^{-11}$ & broad feature\\
28.947  & 2.099 & $\times$ & $10^{-12}$  & --- && --- && \\
31.141  & 8.244 & $\times$ & $10^{-11}$  & --- && --- && \\
---     && --- &                          & 31.379  & $2.099$ & $\times$ & $10^{-12}$ & \\
31.402  & -2.686 & $\times$ & $10^{-11}$  & --- && --- && \\
---     && --- &                         & 31.512  & $6.469$ & $\times$ & $10^{-11}$ & \\
---     && --- &                         & 31.763  & $-7.036$ & $\times$ & $10^{-13}$ & \\
33.332  & -8.177 & $\times$ & $10^{-12}$ & 33.333  & $-5.374$ & $\times$ & $10^{-12}$ & very loud in L1\\
34.826  & 2.099 & $\times$ & $10^{-12}$  & --- && --- && \\
35.762  & 2.306 & $\times$ & $10^{-13}$  & 35.763  & $2.306$ & $\times$ & $10^{-13}$ & known cause (Acromag binary output chassis)\\
37.289  & -3.506 & $\times$ & $10^{-12}$ & --- && --- && \\
37.310  & 1.165  & $\times$ & $10^{-12}$ & --- && --- && \\
---     && --- &                          & 39.763 & $-7.036$ & $\times$ & $10^{-13}$ & \\
42.847  & 2.749 & $\times$ & $10^{-10}$  & --- && --- && \\
---     && --- &                         & 42.918  & $3.967$ & $\times$ & $10^{-12}$ & \\
---     && --- &                         & 43.684  & $-7.036$ & $\times$ & $10^{-13}$ & \\
44.703  & -5.374 & $\times$ & $10^{-12}$ & --- && --- && loud stationary line\\
44.888  & 1.646 & $\times$ & $10^{-10}$  & --- && --- && \\
44.999  & 2.306 & $\times$ & $10^{-13}$  & --- && --- && \\
45.347  & -9.330 & $\times$ & $10^{-10}$ & --- && --- && broad feature \\
46.928  & 2.524  & $\times$ & $10^{-10}$ & --- && --- && \\
46.950  & -2.572 & $\times$ & $10^{-12}$ & --- && --- && \\
---     && --- &                         & 47.683  & $-3.506$ & $\times$ & $10^{-12}$ & \\
48.969  & 1.319 & $\times$ & $10^{-10}$  & --- && --- && broad feature \\
50.250  & -9.111 & $\times$ & $10^{-12}$ & --- && --- && \\
51.009  & 2.095 & $\times$ & $10^{-10}$  & --- && --- && \\
52.617  & 2.078 & $\times$ & $10^{-11}$  & --- && --- && \\
52.807  & 5.178 & $\times$ & $10^{-10}$  & 52.807  & $2.898$ & $\times$ & $10^{-10}$ & \\
53.050  & 1.497 & $\times$ & $10^{-10}$  & --- && --- && \\
53.385  & -1.564 & $\times$ & $10^{-9}$  & --- && --- && broad feature \\
54.724  & 1.172 & $\times$ & $10^{-9}$   & --- && --- && broad feature \\
57.130  & 1.768 & $\times$ & $10^{-10}$  & --- && --- && \\
59.523  & -1.098 & $\times$ & $10^{-11}$ & ---& & --- && \\
59.604  & -7.036 & $\times$ & $10^{-13}$ & ---& & --- && \\
66.666  & -1.191 & $\times$ & $10^{-11}$ & --- && --- && \\
66.757  & 3.506 & $\times$ & $10^{-12}$  & --- && --- && \\
66.877  & 5.836 & $\times$ & $10^{-12}$  & --- && --- && \\
74.505  & -2.572 & $\times$ & $10^{-12}$ & --- && --- && loud stationary line\\
75.033  & 3.272 & $\times$ & $10^{-10}$  & --- && --- && broad feature\\
---     && --- &                         & 83.316  & $6.770$ & $\times$ & $10^{-12}$ & \\
83.445  & 2.306 & $\times$ & $10^{-13}$  & 83.447  & $-1.638$ & $\times$ & $10^{-12}$ & \\
85.830  & -2.572 & $\times$ & $10^{-12}$ & --- && --- && \\
89.406  & -2.572 & $\times$ & $10^{-12}$ & --- && --- && loud stationary line\\
90.040  & 9.178 & $\times$ & $10^{-11}$  & --- && --- && broad, multi-peaked feature \\
96.348  & -1.051 & $\times$ & $10^{-9}$  & --- && --- && no obvious artifact\\
\hline
\hline
\end{tabular}
\caption{Using \textbf{Step 2} of the DM-off search, we identify 53 detector artifacts in
  the Advanced LIGO detectors at frequencies between 20 and 100 Hz. For each artifact,
  we report the $f$ and $\fdot$ values that are associated with the highest $\DMoff2F$
  values in that frequency band if $\DMoff2F > 62$.  Most of these artifacts are weak
  stationary ($\fdot \apprle 10^{-13}$ Hz/s) or wandering lines, but some are
  particularly loud or broad. We find 41 detector artifacts in H1 and
  17 in H1. A few frequency regions contain artifacts in both detectors. One of
  the artifacts (at 35.76~Hz) has a known cause that was independently discovered during an observing
  run.}
\label{table_artifacts}
\end{table*}


\end{document}